# Connectivism as an Innovative Teaching Experience in Initial Physics Teacher Education

Daniel Alejandro Valderrama and
Karen Alexandra Gutiérrez Amaya

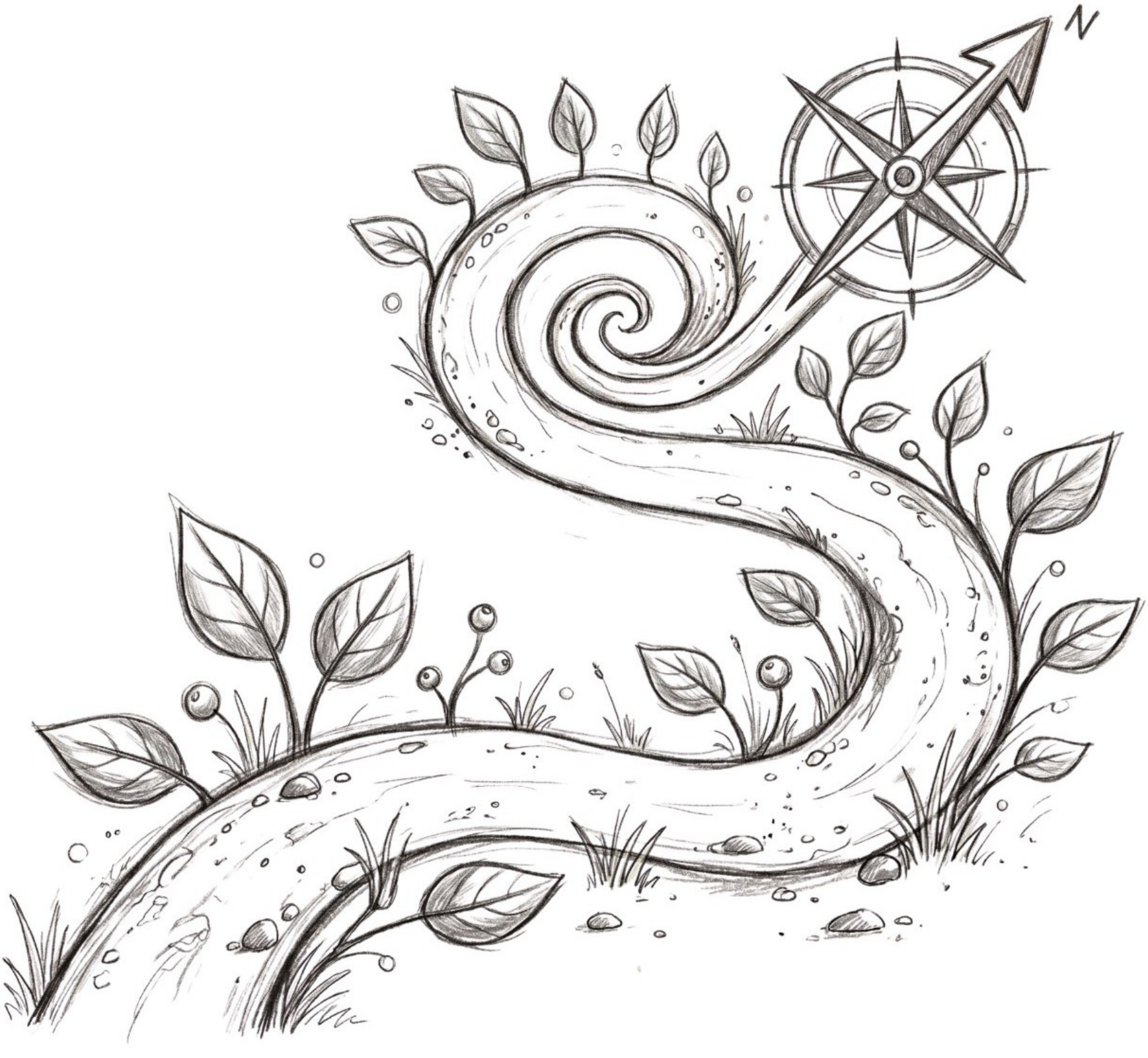

## Introduction

Educational settings have changed considerably under the combined influence of social, cultural, economic and technological developments. Digital technologies have altered educational practices and have required schools, teachers and students to adjust how teaching and learning are organised (BERNATE, 2021). Ready access to information and digital tools has also created opportunities for students to participate more actively in their education. These developments nevertheless pose difficulties, particularly when established educational models are expected to accommodate the conditions of digital environments. Revisiting teaching practices therefore entails considering how new experiences might support the capabilities students need to develop.

These developments have particular implications for science education and for physics teaching. The conceptual demands of physics, together with its frequent presentation as an abstract subject, have historically complicated its teaching and learning. Addressing these difficulties requires approaches that link scientific ideas with students' experiences rather than relying exclusively on the exposition of theoretical content. Science education can provide opportunities for students to identify problems, examine possible explanations and make decisions informed by the circumstances in which they live (RUIZ, 2021).

Such an approach also raises questions about the teacher's role and initial teacher education. Teachers are expected to encourage curiosity, critical examination and problem-solving, while teacher education programmes must consider how digital resources and different accounts of learning can inform classroom practice. The task is to relate technological resources to pedagogical decisions, rather than treating their presence as sufficient evidence of educational improvement.

Connectivism offers a way of understanding learning through the relationships established among people, ideas and technological resources. From this perspective, learning develops through the identification, integration and sharing of knowledge across connected environments (GUTIERREZ, 2012). In initial teacher education, such an approach may create opportunities for digital collaboration, creativity and autonomous learning (BERNATE, 2021). Its relevance to physics teaching lies in examining how students connect conceptual explanations to different sources, shared discussions and practical contexts.

This chapter presents a teaching experience undertaken in Contemporary Physics (CP) classes in the Licenciatura en Ciencias Naturales y Educación Ambiental (LCNEA) programme at the Universidad Pedagógica y Tecnológica de Colombia (UPTC). It explores the use of connectivism in the initial education of prospective physics teachers. The activities sought to establish knowledge networks among students through digital platforms and collaborative spaces, facilitating discussion and the exchange of ideas as part of their preparation for teaching.

**Methodology**

The study involved 21 students enrolled in Contemporary Physics within the LCNEA programme at UPTC. They were organised into four working groups. A qualitative approach informed by connectivism was used to examine the experience. The sequence brought together the selection of resources, discussion of their practical applications, interaction through digital spaces, the preparation of reports and an assessment of the work produced. Figure 1 presents the stages of the implemented process.

Figure 1. Connectivist methodology implemented

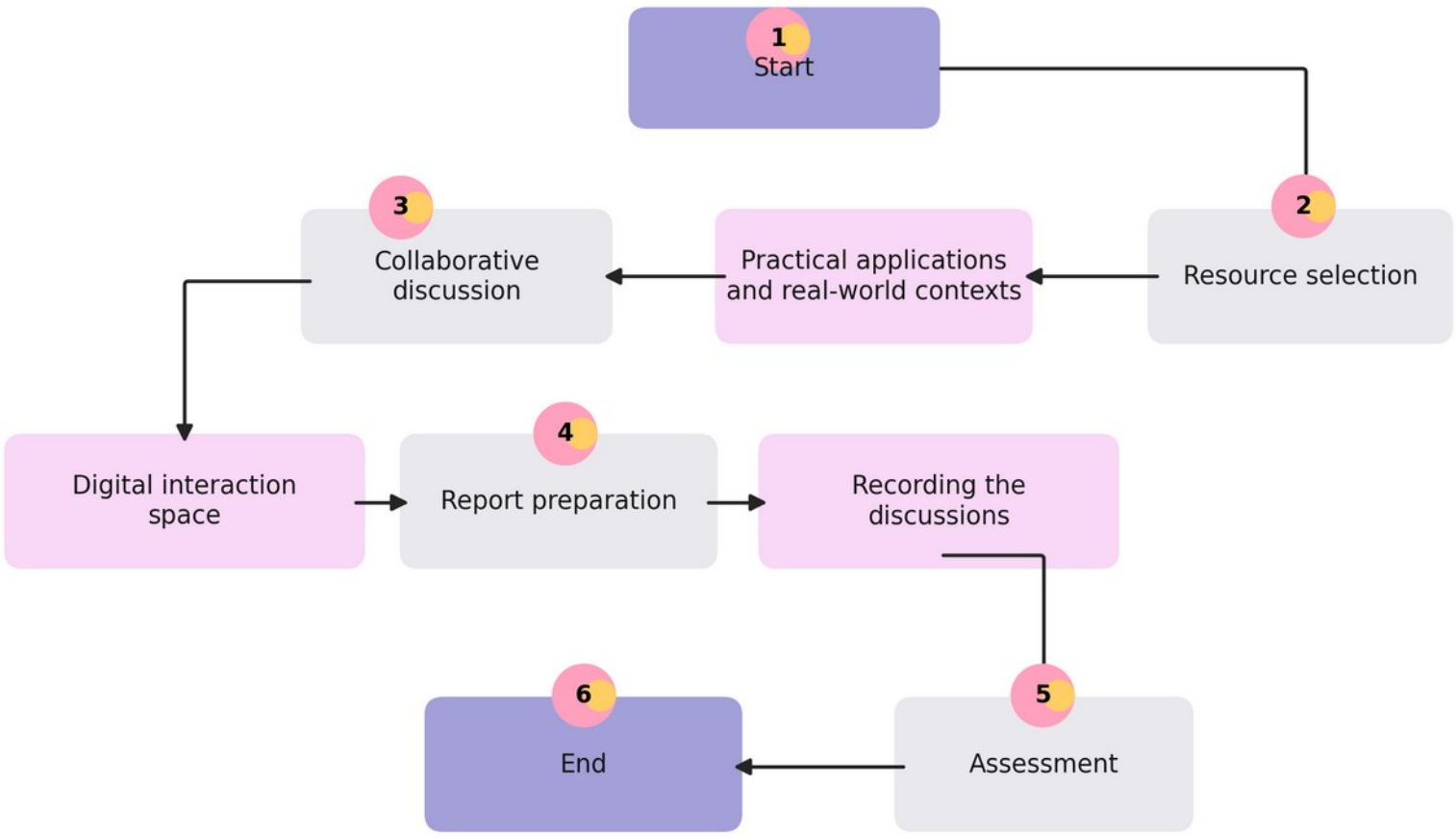

Source: authors' own work.

The group reports were examined through six analytical dimensions, each assessed at three levels. This framework was intended to distinguish different degrees of development in the participants' conceptual and pedagogical work. Table 1 sets out the dimensions, their indicators and the assessment criteria.

Table 1. Dimensions used to analyse the group reports

| Dimension | Indicator | Assessment levels |
|---|---|---|
| Conceptualisation | Identification and understanding of the main concepts. | Superficial: identifies basic concepts without depth.<br>Moderate: connects concepts with simple examples.<br>In-depth: explains concepts accurately using clear and relevant examples. |
| Argumentation | Ability to justify ideas with evidence or practical examples. | Weak: responses based on opinion.<br>Moderate: uses general examples for justification.<br>Strong: provides specific evidence and detailed analysis. |
| Questioning | Formulation of critical questions. | Superficial: simple or general questions.<br>Relevant: specific questions that extend the topic.<br>Critical: questions that open new perspectives or debates. |
| Critical thinking | Analysis of the implications and applications of concepts. | Basic: description without analysis.<br>Analytical: connects concepts with known problems or phenomena.<br>Critical: assesses implications and proposes new approaches. |

| Dimension | Indicator | Assessment levels |
|---|---|---|
| Building connections | Connections among the concepts or phenomena examined. | Isolated: limited connections between concepts.<br>Intermediate: identifies basic connections.<br>Advanced: establishes complex, transdisciplinary connections. |
| Use of digital resources | Selection and use of resources to enrich learning. | Limited: basic use of resources without analysis.<br>Appropriate: suitable use of resources with partial analysis.<br>Integrated: strategic and reflective use of resources for learning. |

Source: authors' own work.

## Results

The experience in Contemporary Physics was organised around learning as the establishment of connections within networks (MONSALVE-LORENTE; AGUASANTA-REGALADO, 2020). Students identified digital materials on physical phenomena and used WhatsApp discussions to exchange interpretations, compare sources and relate course concepts to contributions from classmates and external resources. Each group produced reports recording the subjects addressed, the resources selected and the conclusions reached through discussion. The reports provided the material for examining conceptual understanding, argumentation, questioning, critical thinking, the establishment of relationships and the use of digital resources.

The resulting discussions and reports offer examples of students bringing together information acquired from different sources and relating it to the study of contemporary physics. The following sections describe the work of each group and the assessments presented in the original study.

### A connectivist classroom experience

The first group's reports focused particularly on quantum mechanics, optics and radiation. The students selected material from TikTok and Instagram and also visited educational settings, including the Aerospace Museum and Observatory. These activities connected scientific concepts with practical and technological applications. The combination of online content and face-to-face experiences broadened the sources available for discussion. Table 2 records the topics, applications and resources reported by the group.

Table 2. Reports prepared by Group 1

| Report | Topic and key concepts | Applications | Resource used |
|---|---|---|---|
| 1 | Rayleigh scattering: light, wavelength, visual sensitivity. | Colour of the sky; atmospheric phenomena. | TikTok: Why is the sky blue? |
| 2 | Ionising and non-ionising radiation: ionisation, biological effects. | Health; radiotherapy; security technology. | TikTok: Types of radiation. |
| 3 | Black-body radiation: Planck, Wien and Stefan–Boltzmann laws. | Astrophysics; thermal and solar cameras. | TikTok: Black bodies. |
| 4 | Photoelectric effect: photon energy, quantum nature. | Photovoltaic cells; optical sensors; quantum computing. | Reel: Photoelectric effect. |
| 5 | Schrödinger’s cat: superposition, wavefunction. | Quantum mechanics; foundational principles. | TikTok: Schrödinger’s cat. |
| 6 | Quantum computers: qubits, entanglement. | Algorithms; simulations; emerging technologies. | TikTok: Quantum computers. |
| 7 | Dirac equation: relativity, antimatter. | Quantum computing; cryptography; advanced sensors. | TikTok: Dirac equation. |
| 8 | Quantum well: energy confinement, wave–particle duality. | Semiconductors; optoelectronic devices. | Video: Quantum well. |
| 9 | Tunnelling effect: quantum probability, energy barriers. | Radiotherapy; semiconductors; nuclear processes. | Instagram Reel: Tunnelling effect. |
| 10 | Beta decay: neutrinos, nuclear stability. | Archaeological dating; nuclear physics; medical applications. | YouTube: Beta decay. |
| 11 | Aviation and astronomy: aeronautical innovations, astronomical observation. | Historical communication; technological developments. | Visit: Aerospace Museum and Observatory. |

Source: authors’ own work.

The assessment of Group 1, illustrated in Figure 2, indicates consistently high levels of conceptual understanding across most reports. Students identified the relevant concepts and related them to applications using the digital resources selected. Initial arguments tended to be more superficial, whereas subsequent reports offered more developed justifications through peer interaction and discussion. Questions similarly became more specific and critical over the sequence (TUMINO; BOURNISSEN, 2016). The reports also connected physics with fields such as biotechnology and astronomy, illustrating the possibility of interdisciplinary relationships (RUIZ BERNÉS et al., 2023).

Digital resources played an identifiable role in the group's work. TikTok, YouTube and Instagram provided material for discussion, and the collaborative exchanges made it possible to compare different representations of scientific concepts. The evidence is drawn from the experience and its reported assessments, rather than from a comparison with a non-connectivist teaching group.

Figure 2. Assessment dimensions for Group 1

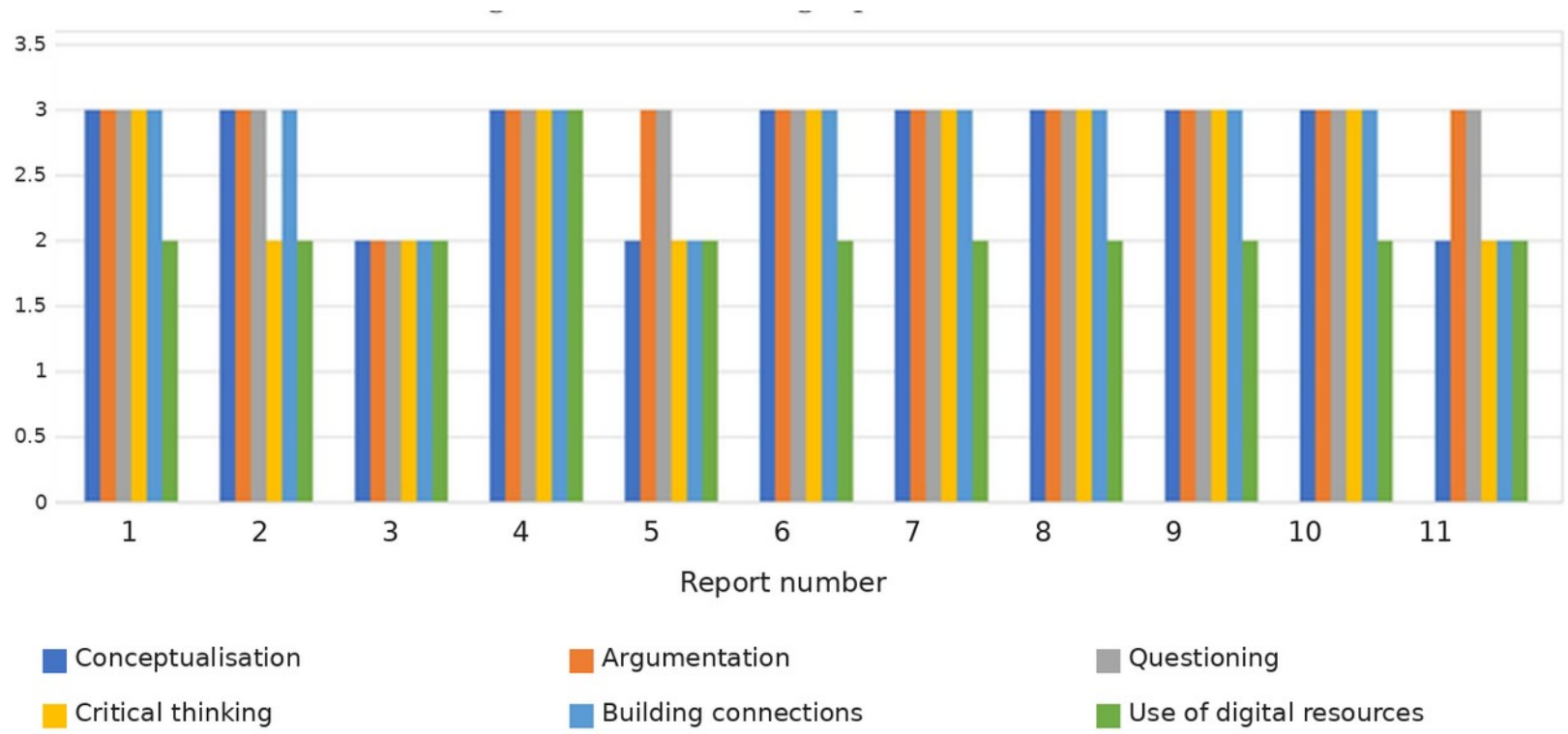

Source: authors' own work.

The second group's reports covered quantum physics, optics, energy, nuclear fusion and climate change. The students drew on explanatory videos and interactive activities to interpret scientific concepts and examine applications, including those associated with sustainable energy. Table 3 lists the subjects and materials they used. The combination of sources is consistent with the emphasis that connectivism places on establishing relationships among different forms of information.

Table 3. Reports prepared by Group 2

| Report | Topic and key concepts | Applications | Resource used |
|---|---|---|---|
| 1 | Refraction of light: focal point, convex lenses, refractive index. | Optics education; lens design; visual entertainment. | Video: The magic arrow. |
| 2 | Climate change: gases, climate modelling. | Climate-change mitigation; renewable energy. | Reflection: Climate change. |

| Repo rt | Topic and key concepts | Applications | Resource used |
|---|---|---|---|
| 3 | Stefan–Boltzmann law: temperature–radiation relationship, black body. | Stellar astronomy; energy optimisation; lighting. | Video: Stefan–Boltzmann law. |
| 4 | Wave–particle duality: wavefunction, photoelectric effect. | Solar panels; diffraction. | Short film: Oppenheimer. |
| 5 | Spectrophotometer: absorbance, monochromator. | Science education; photosynthesis; graphical interpretation. | Video: What is a spectrophotometer? |
| 6 | String theory: fundamental particles, dimensions. | Unification of physical theories; activities with strings. | Video: String theory. |
| 7 | Nature of time: entropy, causality, quantum time. | Philosophical reflection; interdisciplinary teaching. | Video: Grand theory: quantum physics. |
| 8 | Tunnelling and time travel: wormholes, exotic energy. | Theoretical exploration; philosophical analysis. | Video: Is time travel possible? |
| 9 | Nuclear fusion: strong nuclear interaction, extreme conditions. | Sustainable energy; astrophysics; energy crisis. | Video: Nuclear fusion. |
| 10 | Quantum microscopy: photon correlations, microscopic precision. | Medical diagnostics; biotechnology. | Video: Quantum microscope. |
| 11 | Nuclear physics in aviation and astronomy: stellar fusion, fluid dynamics. | Advanced aviation; space exploration. | Visit: Aerospace Museum and Observatory. |

Source: authors' own work.

Figure 3 shows the assessment of Group 2. From the third report, conceptual understanding was generally scored at the upper end of the scale, with students connecting physical concepts to practical applications. The quality of arguments improved towards the end of the sequence. Questioning likewise progressed from less developed enquiries to questions that required more sustained analysis, while critical thinking was evaluated more favourably in later reports (SOPILCA, 2011). The ability to establish connections between physical concepts and other areas, including biotechnology and astronomy, remained consistently high.

Figure 3. Assessment dimensions for Group 2

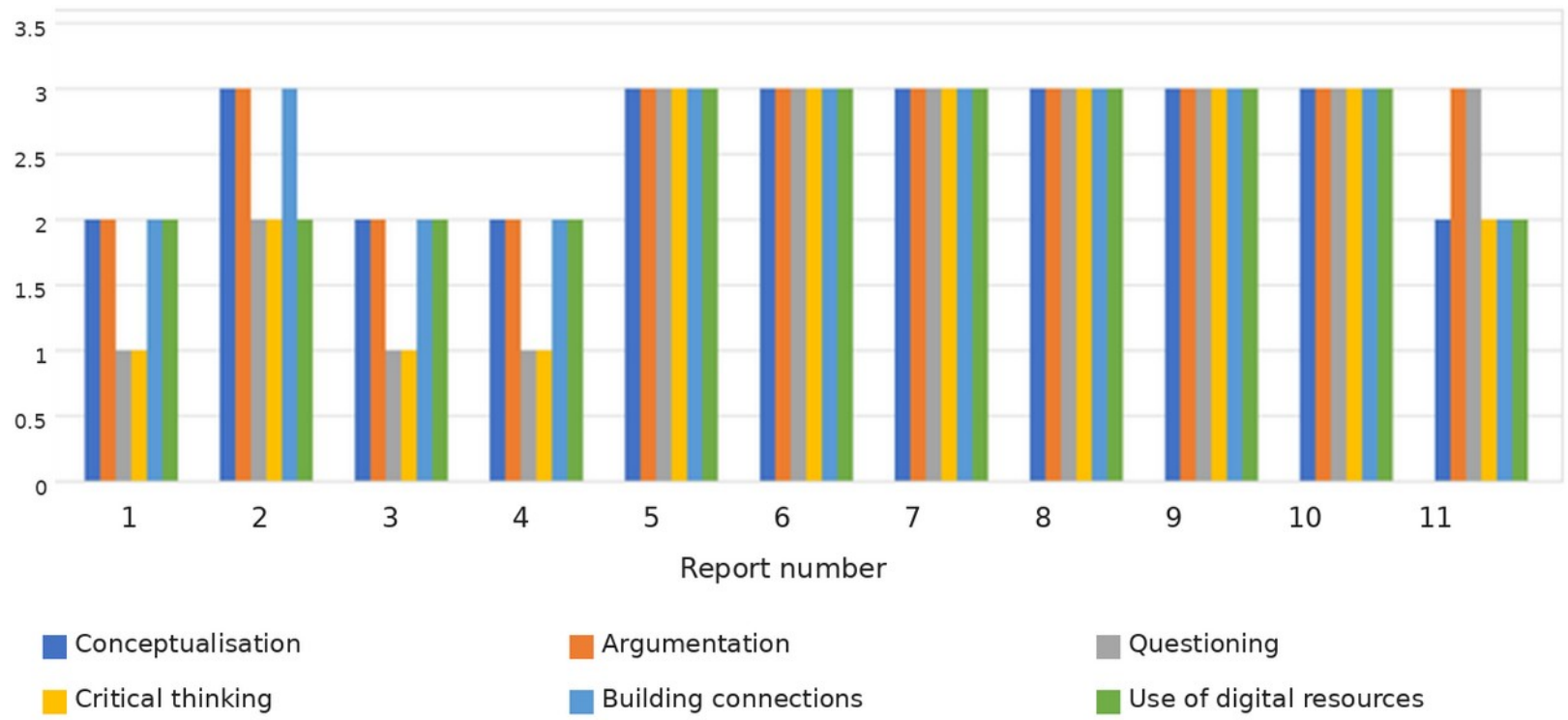

Source: authors' own work.

Group 3 considered thermal conductivity, wave-particle duality, tunnelling and further topics in quantum mechanics. The selected resources provided a starting point for group discussion and collective reflection. Links were drawn with quantum computing, nuclear energy, astrophysics and other technological settings. Table 4 records the contents of the eleven reports and the materials used to support the conversations.

Table 4. Reports prepared by Group 3

| Report | Topic and key concepts | Applications | Resource used |
|---|---|---|---|
| 1 | Thermal conductivity: heat transfer, thermal gradient, insulating and conducting materials. | Electronic systems design; energy efficiency; nanotechnology. | Video. |
| 2 | Decomposition of light: refraction, dispersion, visible spectrum, refractive index. | Optics; spectroscopy; prism and lens design. | Video and group discussion about Newton. |
| 3 | Wave–particle duality: dual nature of light, photons, interference, photoelectric effect. | Quantum computing; cryptography; technological development. | Video and discussion of quantum physics. |
| 4 | Tunnelling effect: probabilistic nature, potential barriers, technological applications. | Semiconductors; radioactivity; nuclear reactions; quantum teaching. | Explanatory video and group discussion. |
| 5 | Use of integrals in physics: area calculations; definite, double and triple integrals. | Mechanics; electromagnetism; thermodynamics; structural design; energy efficiency. | Group discussion and mathematical analysis. |

| Report | Topic and key concepts | Applications | Resource used |
|---|---|---|---|
| 6 | Wavefunction in quantum mechanics: superposition, quantum probability, wavefunction collapse. | Electronics; quantum computing; nuclear energy. | Group discussion and explanatory video. |
| 7 | Uncertainty principle: wave–particle duality, superposition, quantum entanglement. | Quantum computing; molecular diagnostics; philosophical developments. | Video: The uncertainty principle. |
| 8 | Black-body radiation: Stefan–Boltzmann law, Planck distribution. | Thermography; astrophysics; infrared technologies. | Explanatory video and group discussion. |
| 9 | Schrödinger's cat: superposition, wavefunction, Copenhagen interpretation. | Quantum philosophy; interdisciplinary teaching; debates. | Group reflection. |
| 10 | Superposition of white light: interference, coherence, wavelengths, Young's experiment. | Modern optics; human perception; spectroscopy; lighting technologies. | Group discussion and experiments. |
| 11 | Aviation and optics: refraction, wave phenomena, advanced telescope optics. | Science education; astrophotography; spectroscopy; aeronautical development. | Visit to the Aerospace Museum and Observatory. |

Source: authors' own work.

The results for Group 3 (Figure 4) show high assessments of conceptual understanding in most reports, but a reduction in the seventh report. The assessment of argumentation also varied, with a decline in the seventh report and higher scores in the final reports. This variation is relevant when considering how the group used examples and evidence to support its statements.

Questioning remained comparatively consistent, with stronger results from the sixth report onwards. Critical thinking received high assessments, particularly in the fifth report. The chapter interprets these observations in relation to opportunities to reconsider knowledge through repeated engagement with peers and resources (RUIZ BERNÉS et al., 2023). The establishment of relationships was also assessed favourably, as students connected physics concepts to other disciplines (MUFUNGIZI, 2024).

Figure 4. Assessment dimensions for Group 3

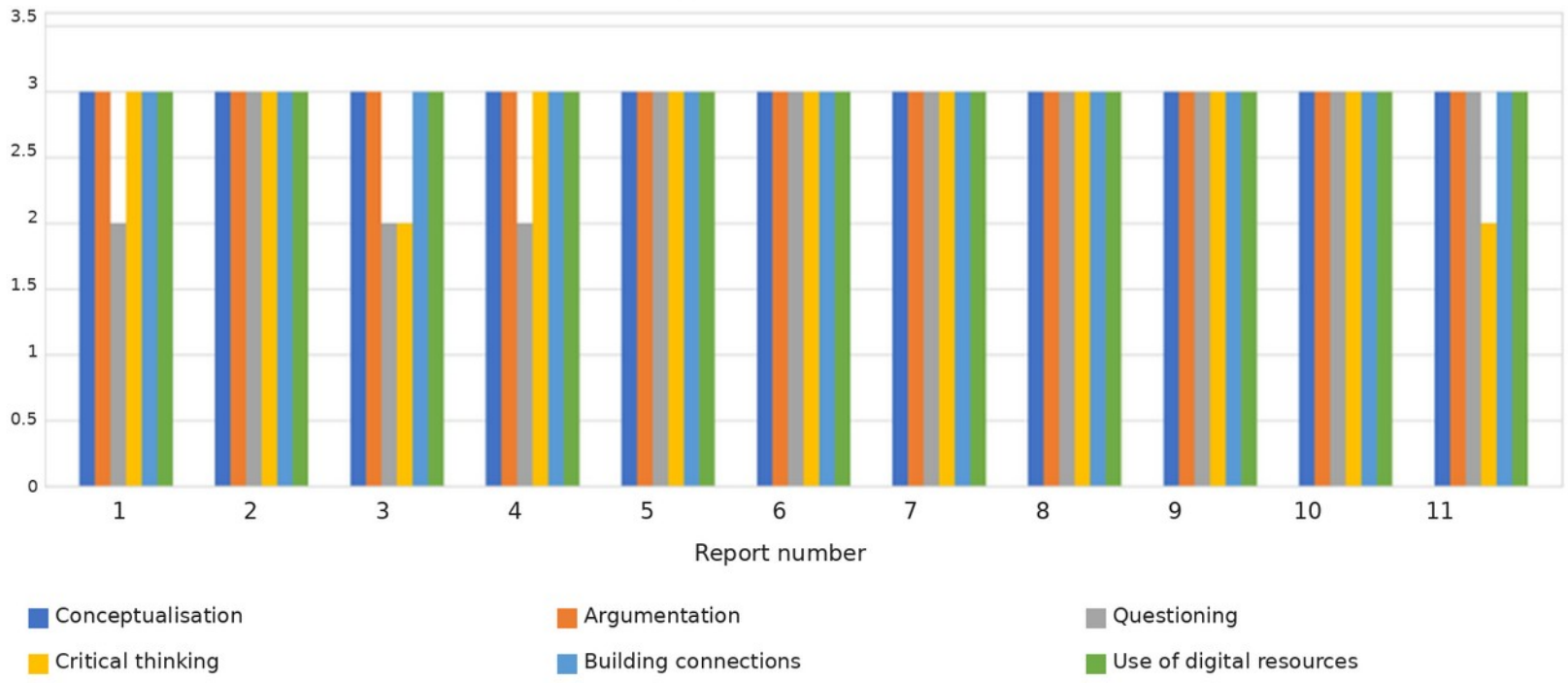

Source: authors' own work.

The fourth group considered subjects ranging from Heisenberg's uncertainty principle to aeronautical technologies. Explanatory videos, film extracts and group discussions were used alongside reflection on interdisciplinary topics. Some reports brought physical concepts into dialogue with questions of neuroscience and wellbeing, extending discussion beyond the immediate course content. Table 5 documents the eleven reports associated with this group.

Table 5. Reports prepared by Group 4

| Report | Topic and key concepts | Applications | Resource used |
|---|---|---|---|
| 1 | Uncertainty principle: conjugate measurements, quantum probabilities, experimental error. | Medical imaging; advanced diagnostics; quantum technologies. | Video: Why is it impossible to measure the speed of light with absolute precision? |
| 2 | Refraction of light: speed, Snell's law, visual distortion. | Lens design; telescopes; optical fibres. | TikTok: The corner of light. |
| 3 | Newton's cradle: conservation and transfer of energy. | Physics teaching; design of educational experiments. | Explanatory video and group discussion. |
| 4 | Quantum physics and wormholes: fluctuations, relativity, space-time. | Theoretical exploration; developments in astrophysics. | Film clip and group discussion. |
| 5 | Time travel: relativity, time dilation, temporal paradoxes. | Interdisciplinary teaching; theoretical research. | Video: Is time travel possible?; discussion. |
| 6 | Stefan–Boltzmann law: thermal radiation, temperature, emitted energy. | Astronomy; materials design; thermal phenomena. | Video: Stefan–Boltzmann law; group discussion. |

| Report | Topic and key concepts | Applications | Resource used |
|---|---|---|---|
| 7 | Energy and wellbeing: wave–particle duality, quantum fluctuations, vacuum energy. | Reflection on wellbeing; neuroscience; interdisciplinary education. | TikTok and group discussion. |
| 8 | Quantum mechanics and the universe: wave–particle duality. | Quantum computing; unified theory; technological developments. | Educational video and group discussion. |
| 9 | Tunnelling effect: quantum probability, energy barriers, uncertainty principle. | Scanning tunnelling microscopy; nuclear fusion; tunnel diodes. | Educational video and group discussion. |
| 10 | Quantum consciousness and multiverse: biocentrism, quantum immortality. | Philosophical reflection; advanced physics teaching. | Educational video and group discussion. |
| 11 | Aeronautical technology and modern physics: aerodynamics, nuclear energy, optics. | Interdisciplinary teaching; technological developments. | Visit: Aviation Museum and National Observatory. |

Source: authors' own work.

For Group 4, Figure 5 presents generally favourable results across the assessment dimensions. Conceptual understanding remained at a high level from the fourth session onwards. Argumentation scores, initially lower, improved in the fifth session. Questions became more developed, and critical thinking received high assessments from the fourth report. The capacity to establish relationships among topics was consistently rated highly, as was the group's use of digital resources. The assessments relate to work within this particular sequence and should be understood in that context.

Figure 5. Assessment dimensions for Group 4

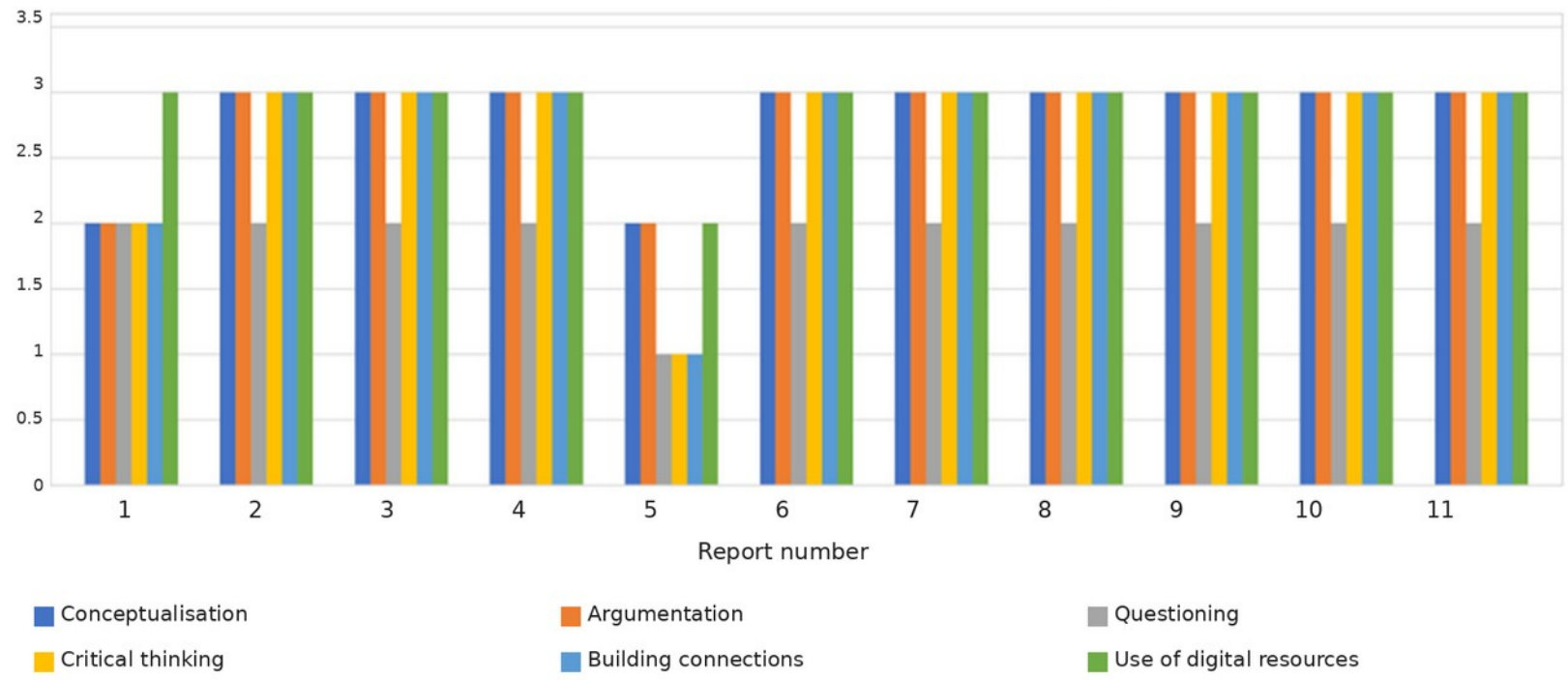

Source: authors' own work.

## Contributions of the experience to physics education, pedagogical preparation and educational innovation

The experience enabled students to explore physics concepts through activities that related the course material to applications and to their preparation as teachers (PRADO, 2021). It also provided occasions for discussion, interdisciplinary connections and considered use of digital technologies. These aspects are situated within a wider debate about connectivism and pedagogical approaches to learning in digital environments (DELGADO-PARRA; SUAREZ-CASTRILLON, 2014). To examine the participants' assessments of the approach, the study collected their responses to statements about collaboration, technological resources and their future teaching.

Table 6 presents students' perceptions of the connectivist activity. The responses were predominantly favourable. The table retains the five original response categories and the number of students selecting each option; each item has 21 responses.

Table 6. Students' perceptions of the connectivist activity (n = 21)

| Statement | Strongly disagree | Disagree | Neutral | Agree | Strongly agree |
|---|---|---|---|---|---|
| Connectivism is a suitable theory for supporting learning in collaborative and digital environments. | 0 | 0 | 1 | 10 | 10 |
| Connectivism facilitates peer interaction and learning in digital activities. | 0 | 0 | 1 | 9 | 11 |
| Using technological resources in learning activities is essential for education in the digital age. | 0 | 1 | 1 | 9 | 10 |
| Weekly selection of resources relating to contemporary physics concepts was useful for my learning. | 0 | 0 | 2 | 2 | 17 |
| Discussions in the WhatsApp group helped me understand the physics concepts studied. | 0 | 0 | 3 | 4 | 14 |
| The rapporteur's consolidation of conclusions clarified the group discussions. | 0 | 2 | 1 | 3 | 15 |
| The activity helped me reflect on using connectivist strategies in my own teaching practice. | 0 | 0 | 1 | 5 | 15 |
| This type of activity helps develop the abilities needed for teaching science in the twenty-first century. | 0 | 0 | 1 | 3 | 17 |
| This methodology enabled me to develop critical and argumentative abilities useful in teacher education. | 0 | 0 | 2 | 6 | 13 |

| Statement | Strongly disagree | Disagree | Neutral | Agree | Strongly agree |
|---|---|---|---|---|---|
| The selected resources seemed appropriate for connecting concepts of modern and contemporary physics. | 0 | 0 | 1 | 4 | 16 |
| Combining conventional and digital resources enriched my learning experience. | 0 | 0 | 0 | 2 | 19 |
| I am motivated to continue using digital platforms for similar activities in the future. | 0 | 0 | 1 | 3 | 17 |

Source: authors' own work.

The majority of students viewed connectivism as suitable for learning in digital and collaborative environments. Most also agreed that it supported interaction among peers and learning through digital activities. These perceptions suggest that participants regarded interaction and networked communication as useful components of the experience (TUMINO; BOURNISSEN, 2016). They should be distinguished from independent measures of learning outcomes.

The use of technological resources was similarly regarded as relevant to education in digital settings. Materials drawn from TikTok and YouTube offered accessible representations of Contemporary Physics topics for the students to discuss. Weekly resource selection was particularly well received: 17 of the 21 participants strongly agreed that it had been useful for their learning. This response provides information about students' judgements of the activity, alongside the descriptions of conceptual work contained in their reports.

The WhatsApp discussions provided a space in which group members could compare interpretations and develop shared accounts of the concepts studied. Fourteen students strongly agreed that those discussions had helped them understand the physical concepts, with four agreeing and three selecting the neutral category. The thematic connections expressed during the activity are represented in Figure 6.

Figure 6. Concepts identified by students during the connectivist activity

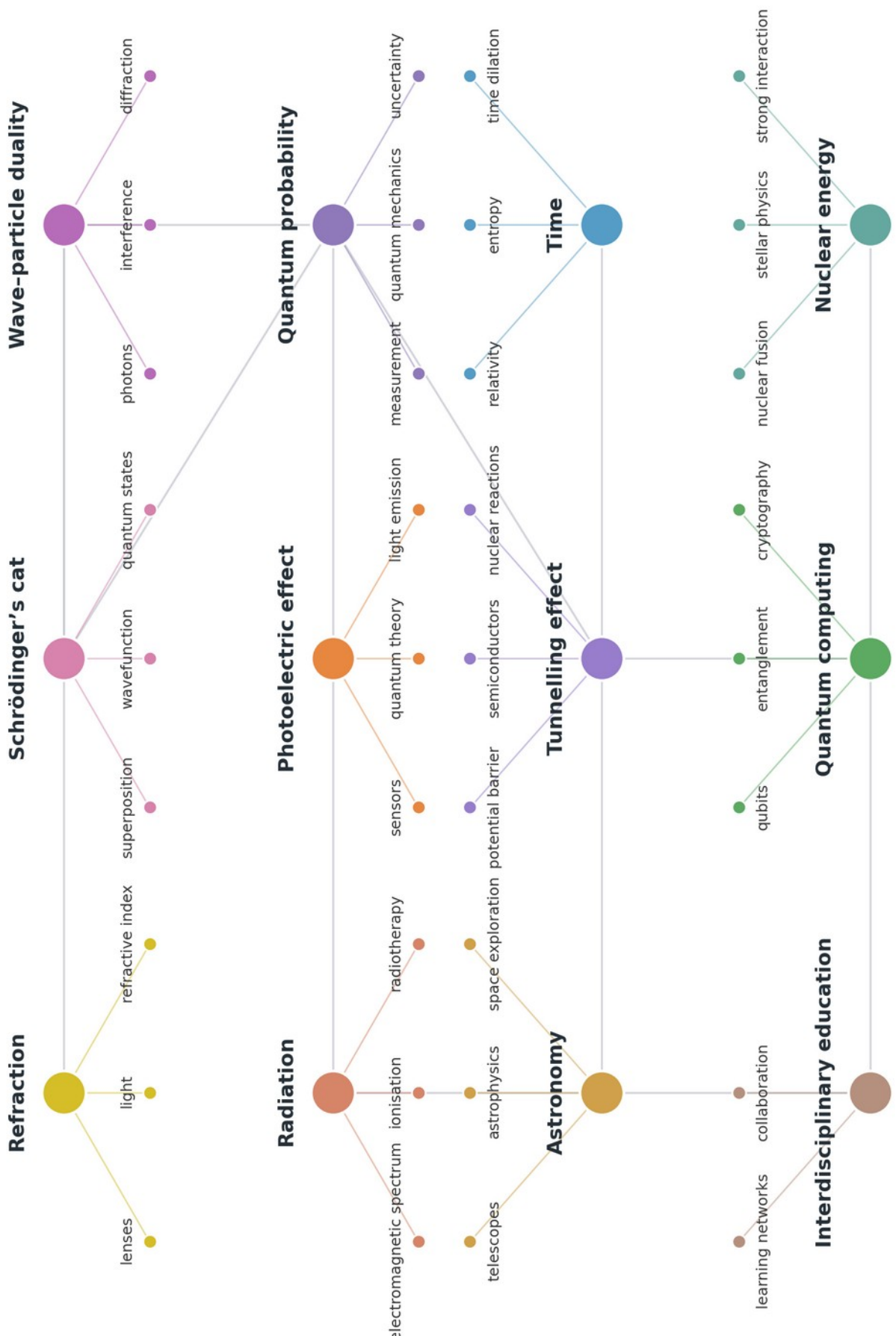

Source: authors’ own work.

Participants also valued the combination of digital and conventional learning resources. Nineteen strongly agreed that the combination had enriched their learning experience and two agreed. Students reported opportunities to practise argumentation, analyse concepts and reflect on their future teaching. The responses describe the participants' perceptions of this specific activity and complement the evidence recorded in their reports.

The preparation of consolidated conclusions by each rapporteur was assessed positively overall, as students associated this stage with clearer group discussions. Participants also reflected on how connectivist strategies could be incorporated into their own teaching. The experience therefore offered an occasion to examine pedagogical decisions as well as the physics content under consideration.

Students recognised the activities as useful for developing argumentation and critical engagement with science topics. They reported relating concepts to other contexts and drawing on different resources in doing so. Their favourable responses to the integration of videos, articles and conventional materials were accompanied by an expressed willingness to continue using digital platforms in similar activities.

## Conclusions

Across the dimensions evaluated, the reports and participant responses showed generally favourable developments during the connectivist teaching sequence. In conceptual terms, students connected topics from Contemporary Physics, including quantum mechanics, optics and thermodynamics, with examples of applications and problems. Working through a network of shared sources provided opportunities to develop and revise these connections collaboratively.

Argumentation changed over the course of the reports. In some groups, initial contributions were relatively superficial, while later justifications drew more directly on evidence and particular examples. Questions likewise moved, in several instances, from general enquiries towards critical questions addressing the topics in greater depth. The group-level figures also show variation between reports, which should not be overlooked when interpreting the overall pattern.

Students established relationships between physics and fields such as biotechnology, astronomy and environmental sciences. This aspect of the activity illustrates the interdisciplinary links considered within a connectivist approach. TikTok, YouTube and WhatsApp supported access to information, discussions about its interpretation and the collective preparation of reports.

The experience indicates how digital technologies and collaborative discussion can be incorporated into initial teacher education in physics. Its contribution lies in the description and assessment of a situated teaching sequence, together with students' accounts of their participation. The qualitative design and absence of a comparison group limit claims about causal effects or transferability to other educational contexts; nevertheless, the findings provide a basis for examining the pedagogical use of networked resources in physics teacher education.